\documentclass[letter,11pt]{article}
\usepackage{jinstpub} 
\usepackage{xspace}
\usepackage{subcaption}
\usepackage{graphicx}
\usepackage{float}

\newcommand{\TZ}{$t_0$\xspace}

\pdfoutput=1 

\usepackage{lineno}
\usepackage{xspace}

\newcommand{\units}[2]{\ensuremath{#1 \; \mathrm{#2} \xspace}} 

\title{Photon detector system timing performance in the DUNE 35-ton prototype liquid argon time projection chamber}

\author[4]{D.~L.~Adams}
\author[24]{T.~Alion}
\author[3]{J.~T.~Anderson}
\author[8]{L.~Bagby}
\author[24]{M.~Baird \footnote{Now at the University of Virginia}}
\author[17]{G.~Barr}
\author[18]{N.~Barros}
\author[8]{K.~Biery}
\author[13]{A.~Blake}
\author[15]{E.~Blaufuss}
\author[6]{T.~Boone}
\author[24]{A.~Booth}
\author[13]{D. ~Brailsford}
\author[6]{N.~Buchanan}
\author[25]{A.~Chatterjee}
\author[21]{M.~Convery}
\author[24]{J.~Davies}
\author[13]{T.~Dealtry}
\author[3]{P.~DeLurgio}
\author[8]{G.~Deuerling}
\author[3]{R.~Dharmapalan}
\author[3]{Z.~Djurcic}
\author[3]{G.~Drake}
\author[21]{B.~Eberly}
\author[8]{J.~Freeman}
\author[18]{S.~Glavin}
\author[9]{R.~A.~Gomes}
\author[3]{M.~C.~Goodman}
\author[21]{M.~Graham}
\author[8]{A.~Hahn}
\author[26]{J.~T.~Haigh}
\author[24]{J.~Hartnell}
\author[11]{A.~Higuera}
\author[8]{A.~Himmel}
\author[14]{J.~Insler\footnote{Now at Drexel University}}
\author[15]{J.~Jacobsen}
\author[8]{T.~Junk}
\author[4]{B.~Kirby}
\author[18]{J.~Klein}
\author[20]{V.~A.~Kudryavtsev}
\author[14]{T.~Kutter}
\author[4]{Y.~Li}
\author[23]{X.~Li}
\author[6]{S.~Lin}
\author[17]{J.~Martin-Albo \footnote{Now at Harvard University}}
\author[20]{N.~McConkey}
\author[1]{C.~A.~Moura}
\author[12]{S.~Mufson}
\author[22]{T.~C.~Nicholls}
\author[13]{J.~Nowak}
\author[3]{M.~Oberling}
\author[8]{J.~Paley}
\author[4]{X.~Qian}
\author[8]{J.~L.~Raaf}
\author[18]{D.~Rivera}
\author[23]{G.~Santucci}
\author[7]{G.~Sinev}
\author[20]{N.~J.~C.~Spooner}
\author[8]{M.~Stancari}
\author[2]{I.~Stancu}
\author[5]{D.~Stefan}
\author[4]{J.~Stewart}
\author[19]{J.~Stock}
\author[8]{T.~Strauss}
\author[16]{R.~Sulej}
\author[10]{Y.~Sun}
\author[20]{M.~Thiesse}
\author[20]{L.~F.~Thompson}
\author[21]{Y.~T.~Tsai}
\author[20]{M.~Wallbank}
\author[20]{T.~K.~Warburton \footnote{Now at Iowa State University}}
\author[6]{D.~Warner}
\author[12]{D.~Whittington \footnote{Now at Syracuse University}}
\author[6]{R.~J.~Wilson}
\author[4]{M.~Worcester}
\author[4]{E.~Worcester}
\author[8]{T.~Yang}
\author[4]{C.~Zhang}

\affiliation[1]{Universidade Federal do ABC, Santo Andr\'e - SP, 09210-580, Brazil}
\affiliation[2]{University of Alabama, Tuscaloosa, AL 35487, USA}
\affiliation[3]{Argonne National Laboratory, Argonne, IL 60439, USA}
\affiliation[4]{Brookhaven National Laboratory, Upton, NY 11973, USA}
\affiliation[5]{CERN, European Organization for Nuclear Research 1211 Geneve 23, Switzerland, CERN}
\affiliation[6]{Colorado State University, Fort Collins, CO 80523, USA}
\affiliation[7]{Duke University, Durham, NC 27708, USA}
\affiliation[8]{Fermi National Accelerator Laboratory, Batavia, IL 60510, USA}
\affiliation[9]{Universidade Federal de Goias, Instituto de Fisica, Goiania, GO 74690-900, Brazil}
\affiliation[10]{University of Hawaii, Honolulu, HI 96822, USA}
\affiliation[11]{University of Houston, Houston, TX 77204, USA}
\affiliation[12]{Indiana University, Bloomington, IN 47405, USA}
\affiliation[13]{Lancaster University, Bailrigg, Lancaster LA1 4YB, United Kingdom}
\affiliation[14]{Louisiana State University, Baton Rouge, LA 70803, USA}
\affiliation[15]{University of Maryland, College Park, MD 20742, USA}
\affiliation[16]{National Centre for Nuclear Research, A. Soltana 7, 05 400 Otwock, Poland}
\affiliation[17]{University of Oxford, Oxford, OX1 3RH, United Kingdom}
\affiliation[18]{University of Pennsylvania, Philadelphia, PA 19104, USA}
\affiliation[19]{South Dakota School of Mines and Technology, Rapid City, SD 57701, USA}
\affiliation[20]{University of Sheffield, Department of Physics and Astronomy, Sheffield S3 7RH, United Kingdom}
\affiliation[21]{SLAC National Acceleratory Laboratory, Menlo Park, CA 94025, USA}
\affiliation[22]{STFC Rutherford Appleton Laboratory, Harwell Campus, Didcot OX11 0QX, United Kingdom}
\affiliation[23]{Stony Brook University,Stony Brook, New York 11794, USA}
\affiliation[24]{University of Sussex, Brighton, BN1 9RH, United Kingdom}
\affiliation[25]{University of Texas (Arlington), Arlington, TX 76019, USA}
\affiliation[26]{University of Warwick, Coventry CV4 7AL, United Kingdom}

\emailAdd{J. Insler (jti27@drexel.edu)}
\emailAdd{A. Himmel (ahimmel@fnal.gov)}
\emailAdd{Z. Djurcic (zdjurcic@anl.gov)}

\abstract{The 35-ton prototype for the Deep Underground Neutrino Experiment far detector was a single-phase liquid argon time projection chamber with an integrated photon detector system, all situated inside a membrane cryostat. The detector took cosmic-ray data for six weeks during the period of February 1, 2016 to March 12, 2016. The performance of the photon detection system was checked with these data. An installed photon detector was demonstrated to measure the arrival times of cosmic-ray muons with a resolution better than \units{32}{ns}, limited by the timing of the trigger system. A measurement of the timing resolution using closely-spaced calibration pulses yielded a resolution of \units{15}{ns} for pulses at a level of 6 photo-electrons.  Scintillation light from cosmic-ray muons was observed to be attenuated with increasing distance with a characteristic length of \units{155 \pm 28}{cm}.}

\collaboration[c]{on behalf of the DUNE collaboration}

\begin{document}
\maketitle
\flushbottom

\section{Introduction}
\label{sec:intro}

The Deep Underground Neutrino Experiment (DUNE) is a dual-site experiment with an international collaboration that will study neutrino oscillation physics, have the ability to observe neutrinos from core-collapse supernovae within our galaxy, study atmospheric and solar neutrinos, and search for nucleon decay.
A beam of neutrinos with 2.5 GeV mean energy will be produced by the Long-Baseline Neutrino Facility (LBNF) at the Fermi National Accelerator Laboratory (Fermilab).  It will be aimed through the Earth's crust toward a 40~kiloton (kt) fiducial mass Liquid Argon Time Projection Chamber (LArTPC) far detector composed of four 10 kt modules situated 1300 km away from the source. This far detector will be located in the Sanford Underground Research Facility in Lead, South Dakota.
The neutrino beam produced at Fermilab will consist almost entirely of $\nu_\mu$.  A near detector situated 574~m from the target will characterize the neutrino beam.   The far detector (FD) will measure $\nu_\mu$ disappearance and $\nu_e$ appearance~\cite{CDRvol2}. The sheer size of the far detector presents multiple engineering challenges: the large cryostat must be cost-effective and maintain the required liquid argon purity, the readout wire planes must be made out of modular components, readout electronics must be chosen that can operate at liquid argon temperatures to minimize noise, and photon detectors must operate in the dedicated field-free regions available in the interior of the detector. In order to prototype and test the necessary technologies and solutions, the DUNE 35-ton prototype detector was constructed and operated at Fermilab.

The DUNE LArTPCs will operate with electron drift times on the order of milliseconds. 
An independent system is therefore required to provide measurements of the time of interaction, \TZ, with resolution equal to or better than \units{1}{\mu s} especially for non-beam interactions such as supernova neutrino scatters, nucleon decay candidates, as well as atmospheric and solar neutrinos interactions~\cite{CDRvol4}. Scintillation light is produced within a few microseconds of the interaction~\cite{LArscintillation} and travels much more quickly than the electrons can drift, so a photon detection system can meet the \TZ requirement. We need \units{1}{\mu s} resolution in order to localize non-beam events in the detector along the drift direction with resolution on par with the TPC wire spacing. The precise location improves energy resolution by allowing corrections for attenuation of the charge signal, which is important to all non-beam physics, and allows fiducialization, which is important for rejecting incoming backgrounds in nucleon decay searches.

The DUNE 35-ton prototype detector was a single-phase LArTPC with an integrated photon detector system built to investigate the DUNE far detector design and components~\cite{CDRvol4}. The experiment was operated in two configurations. 
The first configuration operated between December~20, 2013 and February~15, 2014, and had neither a 
time projection chamber (TPC) nor photon detectors (PD)~\cite{cryostat}. It used dedicated internal purity monitors \cite{puritymonitors} to measure the LAr purity achievable with a non-evacuated membrane cryostat, 
a design envisaged for the DUNE FD.
The cryostat and cryogenic purification system achieved a drift electron lifetime between 2.5 and 3.0~ms, as measured by the purity monitors, which implied an equivalent O$_2$ contamination level of less than 100~ppt. This meets the requirement for LAr purity needed for the DUNE~FD~\cite{CDRvol2}. The second configuration, which collected data between February~1, 2016 and March~12, 2016, included a TPC with
fully integrated photon detectors.  Cosmic rays provided the signals needed to characterize the detector performance because the 35-ton prototype is not in a beamline.  The second configuration's main data-taking period ended when a leak allowed air to contaminate the LAr.  Although subsequent data lacked scintillation light, a useful calibration dataset was obtained via the UV light-based calibration system.


This paper presents a description of the 35-ton prototype detector and the operation of its photon detector system, as well as measurements of the timing performance of the prototype photon detectors and a measurement of the dependence of the strength of scintillation light with the distance between the originating track and the anode plane.

\section{The 35-ton prototype detector}

The 35-ton prototype apparatus was located onsite at Fermilab. It comprised several subsystems, each of which is described below. 
The full apparatus is depicted in Figure~\ref{fig:extmuoncounters}.

\subsection{Membrane cryostat}
The membrane cryostat was made up of a corrugated stainless steel membrane surrounded by multiple layers of insulation and a reinforced concrete structure. The thickness of the insulation was about 0.4~m. The cryostat had inner dimensions of 4.0~m long, 2.7~m wide and 2.7~m high, for a total volume of about 30~m$^3$.  The membrane cryostat was designed by the IHI corporation and constructed at Fermilab under IHI supervision~\cite{cryostat}.

\def\mdsss{5.0in}   
\begin{figure}
\centering 
\includegraphics[width=\mdsss]
{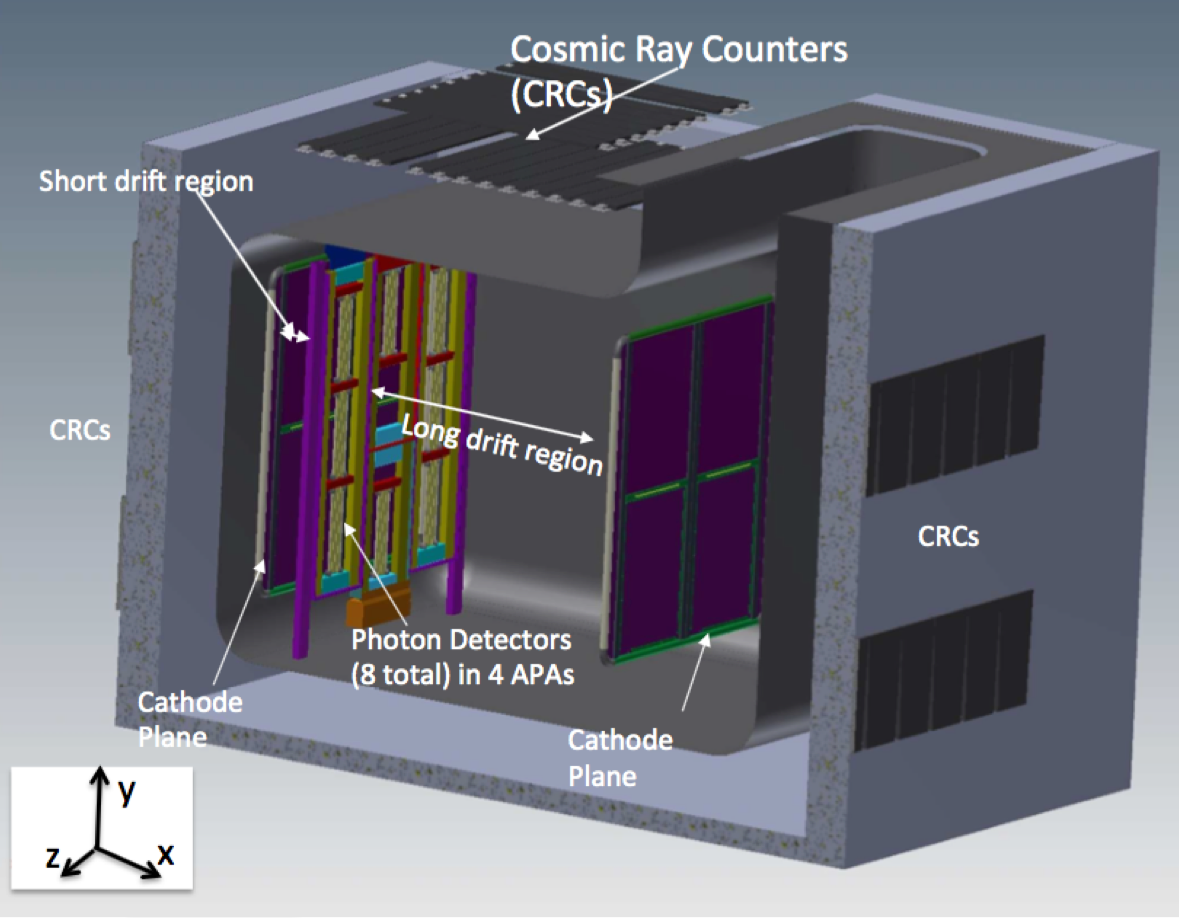}
\caption{\label{fig:extmuoncounters}Drawing of the 35-ton prototype detector, showing positions of Cosmic Ray Counters (CRCs), Photon Detectors, APAs, CPAs and drift volumes. The length of "Long Drift" region is 2.258m and that of the "Short drift" region is 0.302m. 
The APA and CPA planes were each approximately 2 m tall by 1.5 m wide.
CRCs are also present on the near and far walls, which are not shown in the drawing.}
\end{figure}

\subsection{External trigger system}
The external walls of the cryostat were instrumented with 44~pairs of plastic scintillation counters to detect cosmic-ray muons entering and passing through the LAr volume.  These muon counters were developed by the CDF Collaboration and they were used during the Tevatron collider run~\cite{Bromberg:2001gd} before being installed on the 35-ton protoype. The positions of the muon counters are shown in Figure~\ref{fig:extmuoncounters}. In order to record data from throughgoing muons, coincidences between signals from pairs of muon counters on opposite walls of the cryostat were used to trigger detector readout. 

Trigger decisions were made by the Penn Trigger Board (PTB), which was designed and built specifically for DUNE prototypes.  The PTB was equipped with a MicroZed with a Zynq-7Z020 System on Chip (SoC)~\cite{Xilinx}, gigabit Ethernet, 115 I/O ports, one GB of DDR3 RAM and a 33.33 MHz oscillator. It received signals from the muon counters and generated triggers that were sent to the TPC and photon detector readout electronics, described below.  The PTB also could generate triggers internally, both random and periodic.  A triggered readout of the detector is called an ``event'' in this paper.  The PTB provided detailed information on the input trigger signals and counter hit times to the data acquisition (DAQ) system for inclusion in the event data, to be used by the offline analysis. 

\subsection{Time projection chamber}
The TPC 
inside the LAr-filled cryostat consisted of an anode plane and two cathode planes, one on either side of the anode plane.  The anode plane contained four double-sided, modular Anode Plane Assemblies (APAs), and the cathode planes each contained one Cathode Plane Assembly (CPA).  This arrangement formed two drift volumes, one between each cathode plane and the anode plane. High-energy particles passing through the LAr ionized argon atoms along their path, releasing electrons which drifted towards the APAs in the electric fields (with a strength of \units{250}{V/cm} during 35-ton prototype running) maintained between the CPAs and APAs. The APAs had induction wires wrapped around them and vertical collection wires strung across them. When the drifting electrons reached the four APAs, they induced signals on the induction wires before being collected by the vertical collection wires. The induction and collection wire signals were read out by electronics mounted inside the LAr volume and were used to reconstruct the three-dimensional paths of the particles in the drift volumes. The APAs also contained central gaps in which the photon detectors were installed, as described below.
The TPC design and operation is described in greater detail here\cite{35tonTPC}.

\subsection{Photon detector system}
The photon detector system consisted of a collection of photon detectors located inside the APAs, the SiPM Signal Processors (SSPs) used to read out and process the signals from the photon detectors and the calibration system that produced and diffused ultraviolet (UV) light inside the cryostat to monitor the capabilities of the photon detectors.

\def\mdsss{6.0in}   
\begin{figure}[ht]  
\centering 
\includegraphics[width=\mdsss]{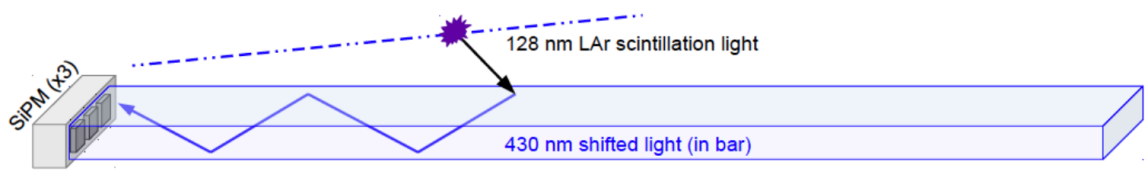} 
\caption{\label{fig:waveguide}Example schematic photon detector panel design, showing the light-guide bar and the SiPMs of the IU or LBNL designs. The light-guide bar has dimensions of \units{50.8}{cm} \ $\times$ \ \units{2.54}{cm} \ $\times$ \ \units{0.6}{cm}.}
\end{figure}

\def\mdsss{6.0in}   
\begin{figure}
\centering 
\includegraphics[width=\mdsss]{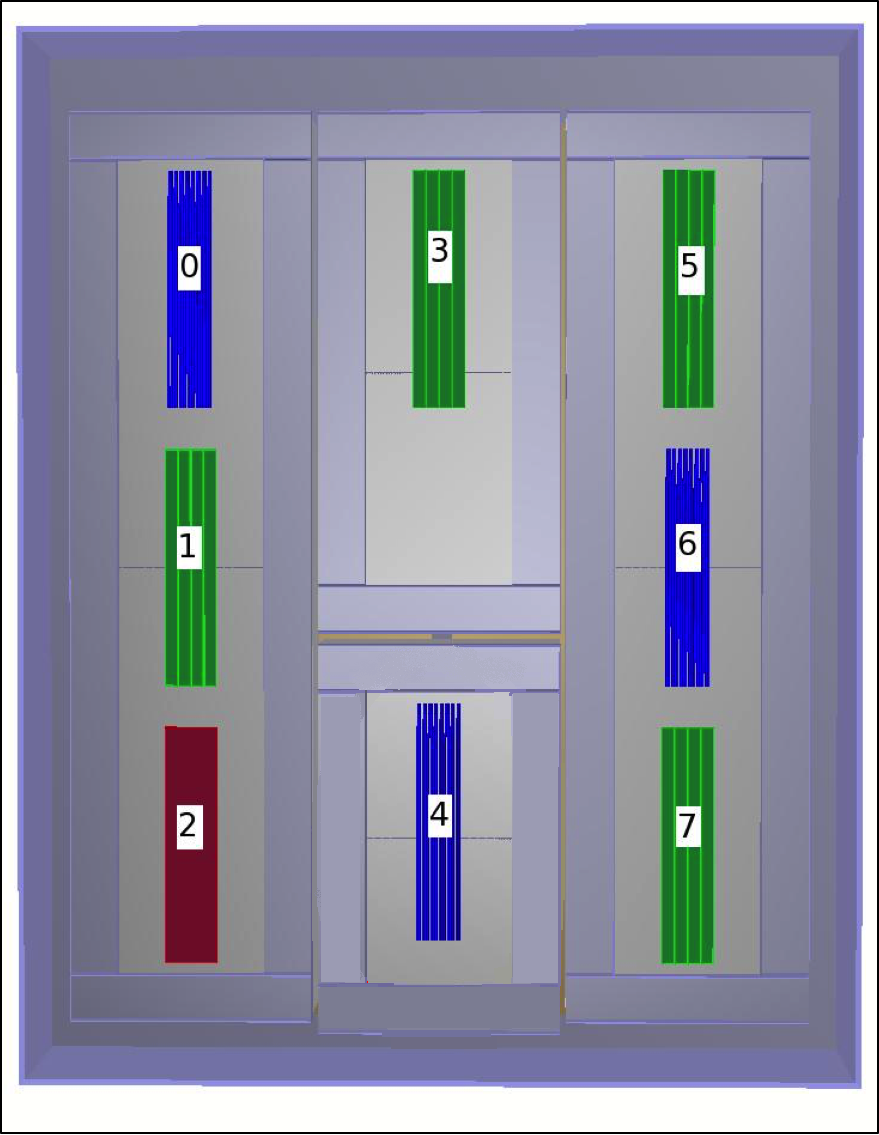} 
\caption{\label{figure:pds}Photon detector positions.} 
\end{figure}
\subsubsection{Photon detector designs}
The 35-ton prototype detector contained eight separate photon detectors installed within the APAs. The photon detectors were constrained in their geometry to fit inside the APAs. The detectors shifted 128~nm scintillation light into the visible range using Tetraphenyl-butadiene (TPB). The shifted light traveled in light guides to silicon photomultipliers (SiPMs) mounted on one or both ends of each detector. For the gain of the SiPM's front-end electronics, we did not amplify SiPM signals but we did use a gain stage with a gain of approximately 10 before the input Analog-to-Digital Converter (ADC), and the amplification was used to adjust the signal amplitude to ADC dynamic range. The SiPMs were run with a bias voltage of 24.5 V.

Figure~\ref{fig:waveguide} shows an example photon detector design, and the photon detector positions are shown in Figure~\ref{figure:pds}.

At the time of the APA construction, several different implementations of this general design were under consideration and were constructed for use in the 35-ton.
However, by the time of operation, some of these designs were considered obsolete, and others had already been tested in standalone systems.
For the 35-ton prototype, the emphasis was on testing the PD system as a whole. Separate measurements of the performance of the different designs  were not performed.

Photon detectors 1, 3, 5 and 7 were assembled at Indiana University (IU) and used a light guide consisting of four acrylic bars produced by IU and MIT. Each bar was attached to three SiPMs, for a total of 12 SiPMs per detector unit. Two light guides were coated with TPB by dipping with a solution of TPB dissolved in dichloromethane at 0.6\% by weight~\cite{MITdipping}; other light guide bars were doped, dip-coated, or flash-heated with either TPB or bis-MSB as described in~\cite{pdsdesignsdune}. The light guides converted scintillation light entering them into photons of visible blue light with wavelength of approximately \units{430}{nm}; total internal reflection caught a portion of the converted photons and conveyed them to the three SensL C-series SiPMs ~\cite{senslC} on the readout end. At a bias voltage of 25.5 V (equivalent to an overvoltage of 5 V above their mean breakdown voltage of 20.5 V), the C-series SiPMs had photon detection efficiency (PDE) at \units{430}{nm} of about 40\%~\cite{Howard:2017dqb}.


Photon detectors 0, 4 and 6 were developed by Colorado State University (CSU). All 3 of the CSU detectors used an array of 32 custom 3 mm square TPB-doped wavelength-shifting (WLS) fibers grouped such that 4 fibers were directed onto the face of a 6 mm square SiPM.  An additional detector was populated with polystyrene bars provided by Lawrence Berkeley National Laboratory and doped with TPB. Due to a cabling error in installation, photon detector 4's readout was not connected; no data were acquired from it during the run.


Photon detector 2 was developed by Louisiana State University and used a TPB-coated acrylic light guide with three S-shaped WLS fibers embedded within it. The TPB converted the scintillation blue light with wavelength approximately \units{430}{nm}, a fraction of which was captured either directly or after reflections off of the internal surfaces of the acrylic fibers, which shifted the light to \units{490}{nm} (green) photons. The fibers guided the green photons to both ends of the detector, where they were read out by SensL B-series SiPMs~\cite{senslB}, one on each end. Since only two SiPMs were required, this design had significantly fewer electronic channels compared to the other designs.

\subsubsection{Photon detector readout}

\begin{figure}[h]
  \centering
\includegraphics[width=15cm]{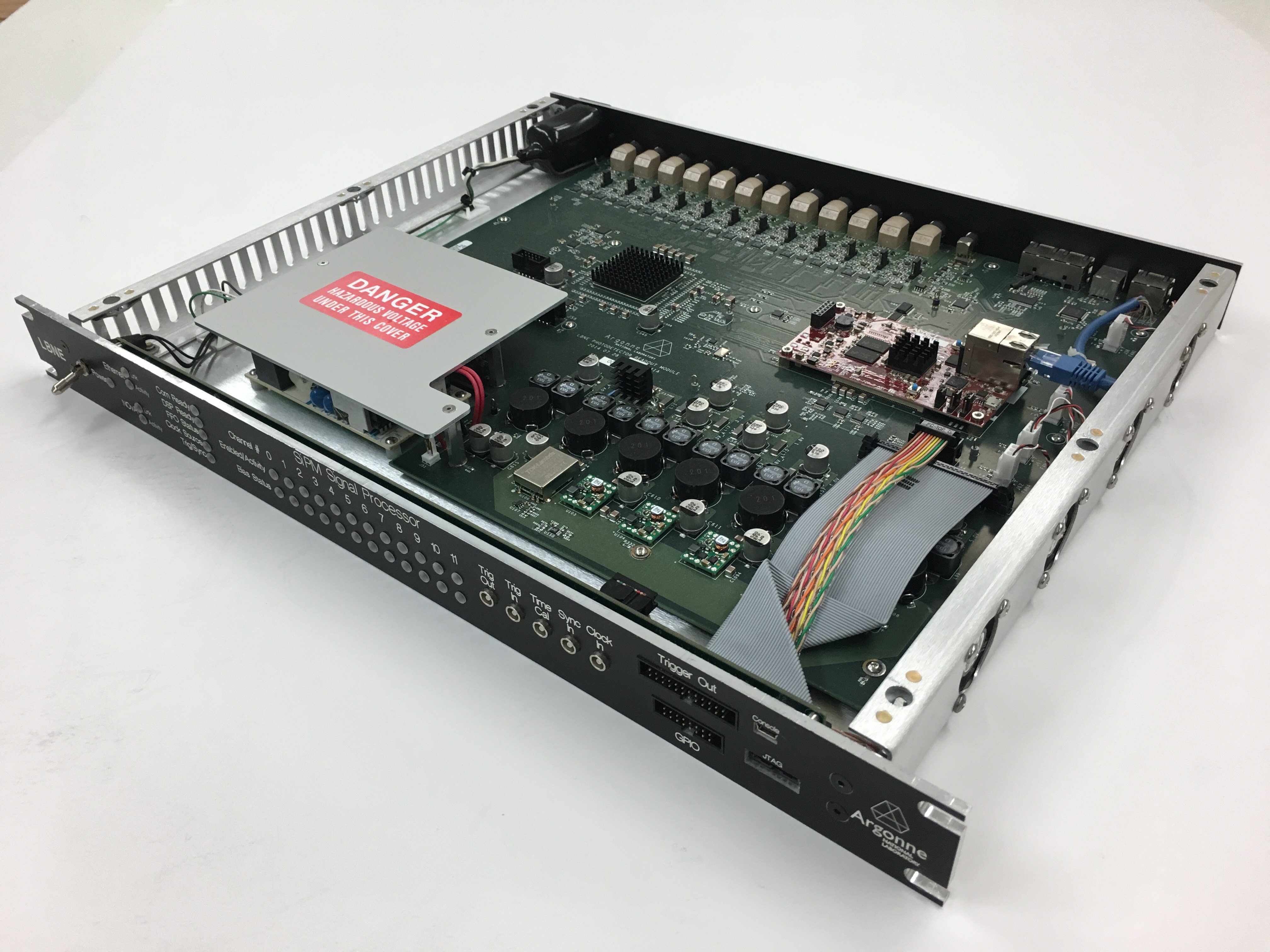}
\caption{A photograph of a SiPM signal processor (SSP).}
\label{fig:PD_ssp}
\end{figure}

Each photon detector's SiPMs were read out by rack-mounted SSP modules, which were designed and constructed by Argonne National Laboratory (ANL). Each SiPM corresponded to one readout channel. The photon detector system made use of seven SSPs in total, connected to the 66 active photon detector channels from all photon detectors except the inactive photon detector~4.  There were no photon detector system front-end electronics in the LAr cold volume. The unamplified signals from the SiPMs were transmitted directly outside the cryostat for processing and digitization, with the advantage that the infrastructure required inside the cryostat was reduced (power, data cables, precision clocks, amplifiers and digitizers). The SSPs received the waveform output from the SiPMs as analog voltages over 20 m long cables. Each SSP included a separate power supply for each channel to provide up to 30~V of bias, controllable in 4096 steps of 7.324~mV, to each SiPM. The module also featured front-end charge injection for performing diagnostics, linearity monitoring and voltage monitoring. A picture of the module is shown in Figure~\ref{fig:PD_ssp}. 

Each SSP channel contained a fully differential voltage amplifier and a 14-bit, 150 megasamples per second ADC that 
digitized the waveforms received from the SiPMs.
In 35-ton running the ADC was run at 128 megasmples per second in order to match the timing of the NOvA system described below. 
The front-end amplifier was configured as fully differential with common-mode 
rejection and received the SiPM signals into a termination resistor that matched the characteristic impedance of the signal cable. 
The processing was pipelined and performed by a Xilinx Artix-7 
Field-Programmable Gate Array (FPGA)~\cite{artix7}.  The digitized data were stored in 
pipelines in the SSP for up to 2048 samples.  
The Artix-7 processed the digitized data from each channel with a leading edge discriminator to detect pulses.
In the standard mode of operation the module performed waveform capture using either an external or an internal trigger. In the latter case, the module self-triggered on a waveform above a predetermined  threshold. 
Because  the Artix-7 FPGA  was programmable  and accessible, it was  possible to  explore  different data processing algorithms  and  techniques. In order for pulses measured in the photon detector to be matched with corresponding readouts of the TPC, the front-end electronics attached a global timestamp from the NOvA timing system~\cite{novatiming} to the data as it was acquired through the SSPs and the TPC electronics.  Quantities from the processed waveforms such as the external timestamp, the peak sum, the baseline sum and the integrated sum were also output in headers.

A Xilinx Zynq FPGA, onboard the MicroZed SoC, handled the slow control (for monitoring and setting parameters) and  data transfer. The SSP had two communication interfaces: USB 2.0 and 10/100/1000 Ethernet. The 1 Gb/s Ethernet supported full TCP/IP protocol and was used to ship the data off the module.

\subsubsection{UV light-based calibration system}

\begin{figure}[h]
\centering
\includegraphics[angle=0,width=12cm,height=9cm]{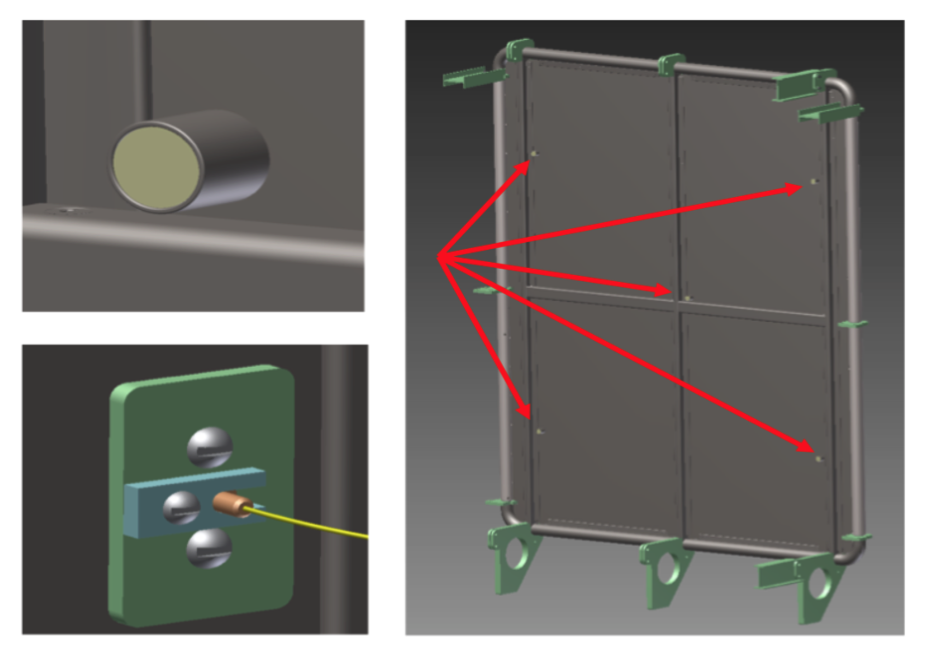}
\caption{ Diffusers (top left-hand figure), which are mounted at five locations shown with arrows (right-hand figure) on the CPA, emit
UV light brought in from the LCM via quartz fibers (lower left-hand figure).}
\label{fig:diffusers}
\end{figure}

The photon detector's calibration system was developed by ANL to verify photon detector dynamic range, SiPM gains and the timing resolution, as well as to evaluate multiple photon detectors' relative efficiencies and to monitor the stability and response of the entire photon detector system as a function of time. It used a set of light-emitting diodes (LEDs) that emitted UV light at a wavelength of 280~nm. Light pulses were generated by a rack-mounted Light Calibration Module (LCM) that contained a set of five LEDs and driver circuitry which utilized an SSP motherboard  for control and communication with the trigger and timing systems. The LCM operated outside of the liquid argon cryostat in a NIM crate alongside the SSPs. Each UV LED in the LCM was coupled to a quartz fiber-optic cable, which transmitted light via an optical feedthrough into the cryostat and the detector volume. 
Each fiber terminated in a diffuser mounted on the CPA in the long drift volume.  The locations of the five diffusers are shown in Figure~\ref{fig:diffusers}. This arrangement improved the uniformity of the illumination of the APA. 
Because the TPB of the photon detectors shifted the wavelength of the 280~nm light similarly to that of the 128~nm LAr scintillation light\cite{gehman},
the detector response should be similar.

\begin{figure}[H]
\centering
 \begin{subfigure}{1.0\textwidth}
   \centering
   \includegraphics[width=0.9\linewidth]{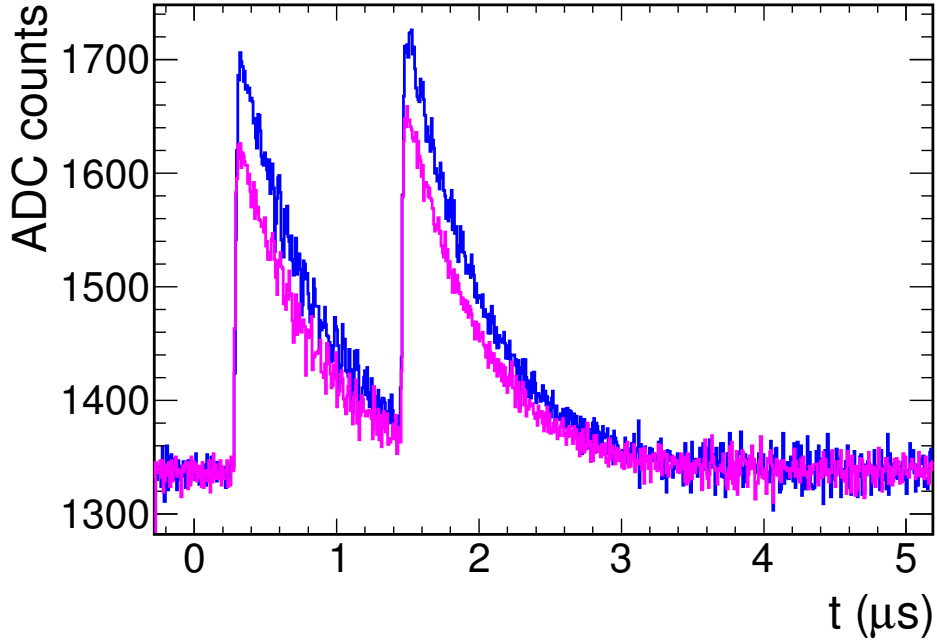} 
   \caption{}
   \label{fig:doublepulse}
   \end{subfigure}
   \begin{subfigure}{1.0\textwidth}
   \centering
   \includegraphics[width=0.88\linewidth]{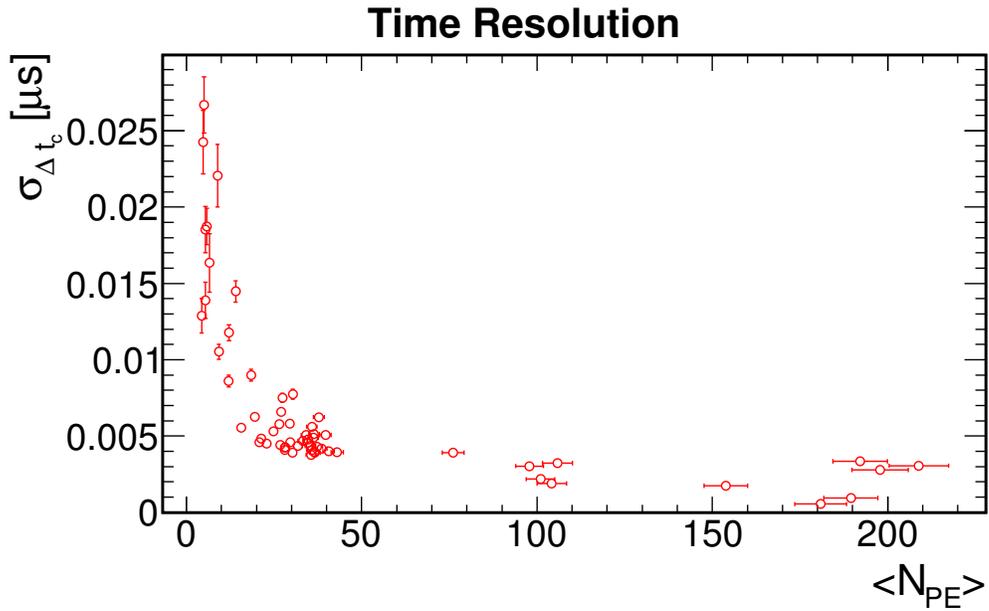}  
   \caption{}
   \label{fig:time_res}
   \end{subfigure}

\caption{(a) Typical example of double-pulse pairs emitted by calibration system and observed by photon detectors. The blue and pink curves are waveforms from two representative channels. (b) Timing resolution of the photon detector system measured with the $\sigma_{\Delta t_C}$ of the distribution of double-pulse time difference pairs of calibration pulses as a function of light intensity (in PE). Horizontal error bars are from counting statistics on the mean number of PE due to different channel efficiencies; vertical error bars are uncertainties in observed time resolution.}
\end{figure}

The calibration system was tested by running the LCM to produce flashes of UV light in the detector volume and provide a trigger to read out the photon detectors and record the response to the light. Special calibration data sets were collected with this setup. The pulse width and the pulse amplitude were varied while flashing only the central diffuser; the other diffusers were also flashed in isolation with constant pulse width and amplitude to test the relative response of the system. Aside from verifying photon detector gains and dynamic range, the calibration system was used in double-pulse modes to measure the timing resolution of the  detector, shown in Figure~\ref{fig:time_res}. 


\section{Photon detector data handling}



The SSP readout could be configured to read out data either upon receiving an external trigger or by self-triggering on the measured waveforms when the amplitude exceeded a threshold defined for each channel. The readout modes could operate together, as well. The SSP output included headers, which contained summary information such as the integrated sum, the peak sum and the baseline sum of the waveform computed by the SSP's on-board signal processing, as well as the full waveform vectors.

Both the externally-triggered mode and the self-triggered mode were employed during data taking. Photon detector data was only written to disk if there was a coincident external trigger, if coincident TPC data was present, or if multiple SSPs had coincident self-triggered waveforms. In the externally-triggered mode, waveforms with the maximum allowed 2048 samples  (a time of about \units{15.5}{\mu s}) were saved when triggers from the muon counters were received. Timing offsets between the muon counters and individual SSPs were found to vary within a small number of PTB ticks (\units{32}{ns}). In the self-triggered mode, the total data rate from the photon detectors needed to be reduced to avoid saturating the maximum data flow capabilities of the DAQ system. Shorter waveforms with 700 samples (about \units{5}{\mu s}) were saved, and an additional filtering software module was added to the DAQ system. The primary contributors to the data rate were radiological contaminants and noise, which could produce signals multiple photo-electrons (PEs) in size, but localized to a single photon detector. The rate could thus be significantly reduced by requiring coincidences in multiple photon detectors. Because all of the channels for each photon detector were read out by a single SSP, this could be achieved by requiring coincident hits in multiple SSPs. Even with these mitigations, high noise in the integrated system required the photon detector readout threshold to be set to at least 2.3~PE.


\section{Photon detector performance}


The performance of the prototype photon detector system was studied using two datasets:  cosmic-ray muons penetrating the detector from the atmosphere and UV light pulses from the calibration system mounted inside the cryostat. Both datasets were used to study timing resolution, and the cosmic-ray dataset was used to measure the photon detector response vs. position in the detector. LAr scintillation light, such as that produced by the cosmic-ray muons entering the cryostat, is known to contain a ``prompt'' component with a decay time constant less than \units{6.2}{ns} and a ``late'' component with decay time constant around \units{1.3}{\mu s}~\cite{LArscintillation}. The overall response function of our photon detector system has a \units{1}{\mu s} time scale that is close to the time constant of the late light, meaning that a large fraction of the "late" light is actually overlapping with the prompt signal. Therefore our photon detector system typically observed a mix of the prompt and the early part of the late light with some fraction of the late part of the late light from the cosmic-ray muons.

\subsection{Optical hit finding}

The first step in the offline analysis of these data was the creation of "optical hits", which were found by identifying peaks in the recorded waveforms. An example waveform is shown in Figure~\ref{fig:waveform}.
Information stored for each hit included the time, the width, the area and the amplitude.
In order to find optical hits, the waveform was first processed by a pedestal-finding algorithm, 
which used the levels before and after the pulse (itself identified by a rapid rise in amplitude of the pulse) to calculate the pedestal.

The optical-hit-finder algorithm first subtracted this pedestal from each sample of the waveform.
Then it looped through all of the samples in the waveform and searched for a value that exceeds a "primary" threshold.
Once found, the algorithm defined a region of interest around that peak as the range during which the ADC values exceed a "secondary" threshold. The primary and secondary thresholds were fixed parameters. If the width of the region of interest was sufficiently wide, the optical hit was created and assigned the time of the peak value sample within the hit. 

\subsection{Timing resolution measurement}
\label{sec:timingres}
In order to measure the timing resolution of the photon detector system with respect to the muon counters, the time difference ($\Delta t_\mu$) between the reconstructed optical hit times from the photon detectors and the trigger time constructed by the PTB from the coincidence of two muon counter signals was computed. All runs containing photon detector data externally triggered by the muon counters as our data sample were used. To exclude the large noise present in the data, only hits with a peak of greater than an ADC count of about 100 above pedestal, corresponding to approximately 5.2~PE, were used. Most of the photon detectors displayed run-to-run fluctuations in the timing offset between the muon counters and photon detector signal processing that were not inherent to the photon detectors themselves, so we chose to look at signals from photon detector 3, which had the most consistent time offset between the photon detector and muon systems over the full run. Looking at $\Delta t_\mu$, a narrow peak of width $<$ \units{100}{ns} was apparent.
\begin{figure}[H]  
\centering 
   \begin{subfigure}{1.0\textwidth}
   \centering
 \includegraphics[width=0.9\linewidth]{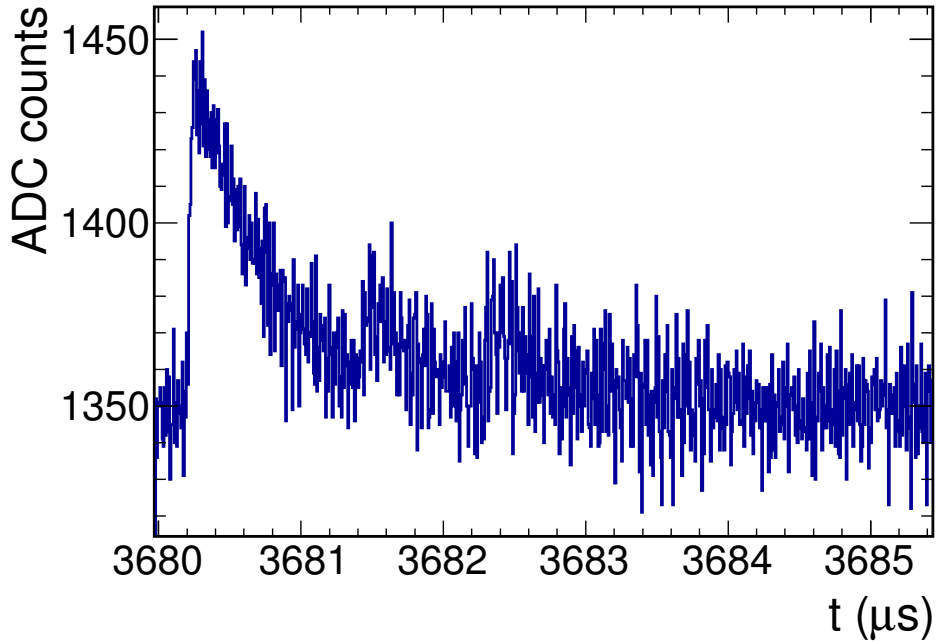} 
   \caption{}
   \label{fig:waveform}
   \end{subfigure}
   \begin{subfigure}{1.0\textwidth}
   \centering
   \includegraphics[width=0.9\linewidth]{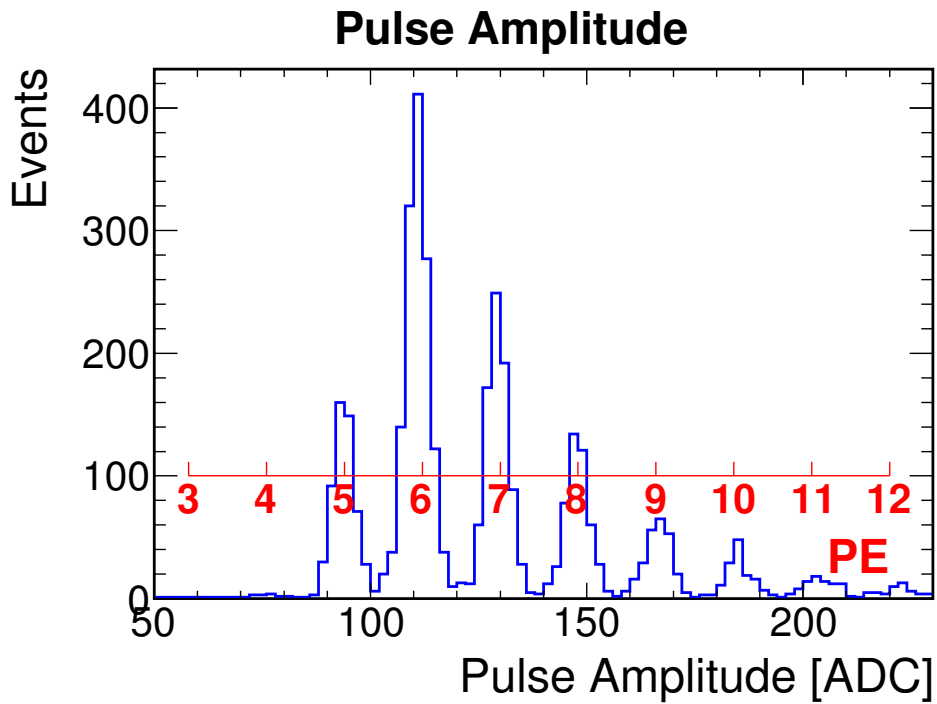} 
   \caption{}
   \label{fig:PE_peaks}
   \end{subfigure}
   \caption{(a) Waveform from a photon detector showing a clear optical hit approximately 4~PE in size. This waveform was externally-triggered by the muon system to collect 700 samples  (about \units{5.4}{\mu s} in length each).\\(b) PE distribution from a typical photon detector readout channel.}
\end{figure}

 In single runs, all times showed up within a single bin, indicating that the time resolution of the photon detector is better than the PTB tick length of \units{31.25}{ns}. This establishes that the resolution surpasses the requirement for DUNE; by comparison the readout for the TPC is digitized with a period of typically \units{500}{ns}. In fact, this is a measurement of the combined timing resolution of the PD system, the muon counters, and the timing system, which was the limiting factor. This measurement was significant because it set an upper limit on the timing resolution of the PDs using naturally occurring cosmics rather than an internal calibration system.


To further quantify the timing resolutionof the photon detectors themselves, a measurement using the in situ calibration system was performed. Data were taken after the primary data run with the TPC powered down and with a high level of nitrogen present in the LAr, which effectively extinguished any scintillation light. Figure~\ref{fig:PE_peaks} displays the pulse amplitudes of a typical SSP channel. Distinct peaks show a clear separation corresponding to different numbers of PE. Assuming linearity, SiPM gain was found to be approximately 18 ADC counts per single PE. With all the SiPMs operated at the same bias voltage the gain had a variance of about 5\% among the channels. Calibration data were collected in the self-triggering mode and the hit-finding thresholds used in the standard analysis were ignored; instead, a threshold of 80 ADC counts above pedestal (approximately 4.5~PE) was applied. In our timing resolution studies, pulses were required to have $N_{\rm{PE}} \ \geq \ ~6$. 

For these runs, the LCM was configured to emit two pulses of the same duration, the second of which was emitted with a well-defined time delay relative to the first. An example of two readout waveforms taken in this double-pulse mode is shown in Figure~\ref{fig:doublepulse}. In this example, both pulses were \units{3.9}{ns} long and the time difference between emitted pulses was \units{1172}{ns}. The jitter on this \units{1172}{ns} interval was measured to confirm that it was negligible; for all calibration configurations (including various pulse widths and pulse-to-pulse delays), the jitter between two light pulses was been measured to be less than \units{300}{ps} FWHM.

For each SiPM channel a few hundred double-pulse events were collected. Then the time difference $\Delta t_C = t_2 - t_1$ between two pulses was measured as a function of the number of PE observed at a given SiPM. For this study, the pulse time was taken as the time at two thirds of the maximum pulse height at the leading edges of the first and second pulses. For each SiPM channel, the mean number of PE, $\langle N_{\rm{PE}} \rangle$, and the resolution (standard deviation) of $\Delta t_C = t_2 - t_1$, called $\sigma_{\Delta t_C}$ was calculated. Figure~\ref{fig:time_res} shows $\sigma_{\Delta t_C}$ as a function of $\langle N_{\rm{PE}} \rangle$ for the 46 channels that were used for this study; these resolutions are a measure of the difference between the two leading edges of pulses detected by the SiPMs and readout electronics for all functioning photon detector channels, as is shown in Figure~\ref{fig:doublepulse}. Twenty channels were excluded due to high noise or other problems. The light-pulser emitted photons defined by pulse widths of \units{3.9}{ns} and \units{11.7}{ns}; the points in Figure~\ref{fig:time_res} with $\langle N_{\rm{PE}} \rangle$ between 5 and 25 were measured with \units{3.9}{ns} wide pulses, and the points with $\langle N_{\rm{PE}} \rangle$ greater than 25 were measured with \units{11.7}{ns} wide pulses. The larger measured time resolutions for small $\langle N_{\rm{PE}} \rangle$ were due to multiple factors, including variations due to the times that photons were emitted within their respective pulses, propagation delays, multiple paths and reflections, variations in TPB fluorescence emission time, and variations in electronic noise.  For larger light pulses the aggregate behavior significantly reduced these variations. The overall time resolution dependence on $\langle N_{\rm{PE}} \rangle$ in Figure~\ref{fig:time_res} followed $\sim \frac{1}{\sqrt{\langle N_{\rm{PE}}} \rangle}$ behaviour.

The time resolution was approximately \units{4}{ns} for 50~PE pulses and worsened to approximately \units{15}{ns} for 6~PE pulses; the approximate \units{4}{ns} resolution was due to the leading edge discrimination using a 128 MHz sampling rate with \units{7.8}{ns} period. The time difference of \units{1172}{ns} between two pulses was due to the leading edge of the SiPM pulse falling in between two ticks \units{7.8}{ns} apart and therefore having an average of about \units{4}{ns}.
 This resolution is better than the global timing resolution found with muons tagged by the external triggers in Section ~\ref{sec:timingres} because the photon detector system has a faster internal clock than the PTB. 




\subsection{Photon detector response vs. position}

The amplitude of the observed light as a function of the distance between the track producing it and the anode plane was also studied in order to quantify the attenuation of scintillation light due to scattering, absorption, and geometrical effects in the 35-ton prototype detector. From the full data set of externally triggered events, events with muons that triggered one of the seven pairs of counters located directly opposite each other on the cryostat walls were selected. 
These are tracks that are traveling roughly parallel to the $z$-axis of Figure \ref{fig:extmuoncounters} at varying $x$ positions. Events were required to have no more than the two triggers from the muon counter pairs. The reconstructed optical hits were required to have an amplitude above a threshold of 100~ADC counts (approximately 5.2~PE) and were required to not be from one of the 10 noisy optical channels. 

\begin{figure}
\centering 
\includegraphics[width=5.5in]{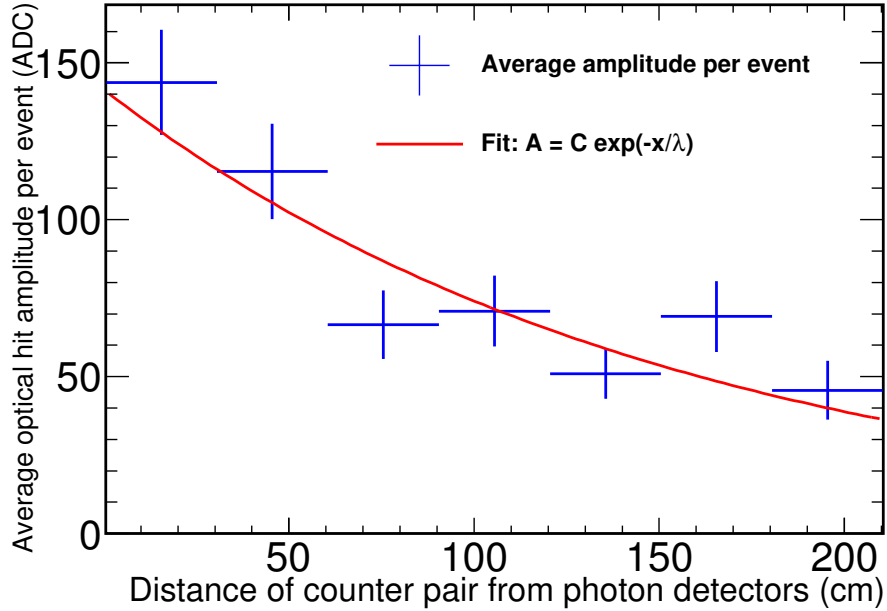}
\caption{\label{fig:average_amplitude_per_event} Average optical hit amplitude per event vs. counter pair positions, fit with $A = C e^{-x/\lambda}$ where $A$ is avg. amplitude, $\lambda$ is the characteristic length, $x$ is distance of the tracks from PDs and $C$ is a normalization constant. Error bars are statistical errors on mean hit amplitudes per bin.}
\end{figure}

For each event that triggered a pair of muon counters, all reconstructed amplitudes of optical hits were summed; if no optical hit was present, the summed amplitude was set to zero. Events with no optical hits above the threshold of 5.2~PE were treated as zeros and not excluded from the average to avoid biasing it towards events with brighter scintillation light. The summed amplitudes over all of the events with triggers were averaged, yielding an average measured light yield per event. Figure~\ref{fig:average_amplitude_per_event} shows the measured averages along with an exponential fit.  The characteristic length in this fit is \units{155 \pm 28}{cm} (statistical error only).  This length includes geometrical effects, scattering, absorption on the field cage and CPA walls, as well as absorption in the liquid argon; no attempt has been made to measure the impacts of these effects separately. Since this measurement includes the above factors, it should not be directly compared to measurements of the inherent attenuation length in LAr. The measured characteristic length of 155$\pm$28 cm is greater than the typical values of 50 to 90 cm for the scattering length of scintillation photons in LAr due to Rayleigh scattering~\cite{indexofrefraction} because our measurement covers a large area by summing over seven photon detectors. Scintillation light is only lost when it scatters into the walls of the cryostat, and scattering tends to move the scintillation light to a different detector rather than the light being lost. Due to this larger coverage, the attenuation length is longer than the typical scattering length.



\section{Conclusion}

The data run of the 35-ton prototype detector demonstrated the use of the DUNE photon detector designs integrated into a LArTPC detector. Based on a data sample of all runs externally triggered by the muon counters containing photon detector data, the timing resolution was found to be better than \units{32}{ns} with measured muons for a single photon detector.
The calibration system with UV diffusers was also tested and used in specialized runs such as the timing resolution measurement. A time resolution of \units{15}{ns} was found for pulses at a level of 6~PE from double-pulse calibration measurements.
The photon detector system therefore exceeds the timing resolution requirements of the DUNE far detector, demonstrating a timing resolution significantly better than \units{1}{\mu s}. A characteristic length (including scattering, effects of the geometry, scattering, and absorption on the field cage and CPA walls as well as absorption in the liquid argon) for scintillation light of \units{155 \pm 28}{cm} (statistical error only) was found based on a sample of through-going muons as tagged by the external muon counters. 

\acknowledgments
This material is based upon work supported in part by the following: the U.S. Department of Energy, Office of Science, Offices of High Energy Physics and Nuclear Physics; the U.S. National Science Foundation; the Science and Technology Facilities Council of the United Kingdom, including Grant Ref: ST/M002667/1; and the CNPq of Brazil. Fermilab is operated by Fermi Research Alliance, LLC under Contract No. DE-AC02-07CH11359 with the United States Department of Energy.
We would like to thank the Fermilab technical staff for their excellent support and the MIT Neutrino Group for supplying acrylic bar lightguides.


\end{document}